\documentclass[twoside,fleqn]{article}
\usepackage{epsfig}
\usepackage{espcrc2}
\newcommand{\bc} {\begin{center}}
\newcommand{\ec} {\end{center}}
\newcommand{\bqa}{\begin{eqnarray}}
\newcommand{\eqa}{\end{eqnarray}}
\newcommand{\nn}{\nonumber}

\newcommand{\ga}{\gamma}

\newcommand{\si}{\sigma}
\newcommand{\om}{\omega}
\newcommand{\ta}{\tau}
\newcommand{\ka}{\kappa}
\title{\vskip -100pt
\mbox{} \hfill BI-TP 2001/24\\
\mbox{} \hfill DESY 01-153\\
\mbox{} \hfill October 2001\\
\vskip 45pt
Meson Spectral Functions at finite Temperature}
\author{I.~Wetzorke\thanks{Present address: NIC/DESY Zeuthen,
Platanenallee 6, D-15738 Zeuthen, Germany}
with F.~Karsch, E.~Laermann, P.~Petreczky and S.~Stickan
\thanks{The work has been supported by the TMR network ERBFMRX-CT-970122
and the DFG under grant FOR 339/1-2.}
\\[3mm]
Fakult\"at f\"ur Physik, Universit\"at Bielefeld, D-33615 Bielefeld,
Germany}
\begin{document}
\begin{abstract}
The Maximum Entropy Method provides a Bayesian approach to reconstruct
the spectral functions from discrete points in Euclidean time.
The applicability of the approach at finite temperature is probed with the
thermal meson correlation function. Furthermore the influence of
fuzzing/smearing techniques on the spectral shape is investigated.
We present first results for meson spectral functions at several
temperatures below and above $T_c$. The correlation functions were
obtained from quenched calculations with Clover fermions on large
isotropic lattices of the size $(24-64)^3 \times 16$.  We compare the
resulting pole masses with the ones obtained from standard
2-exponential fits of spatial and temporal correlation functions
at finite temperature and in the vacuum. The deviation of the
meson spectral functions from free spectral functions is examined
above the critical temperature.
\end{abstract}
\maketitle
\section{INTRODUCTION}
The maximum entropy method (MEM) \cite{Jar96} has been applied successfully
to reconstruct hadronic spectral functions from correlation functions in
Euclidean time calculated at zero temperature in quenched QCD
\cite{Asa00,Yam01}. It was demonstrated that the spectral shape reflects
ground and excited state contributions, which yields reliable results for the
respective masses and decay constants.

At finite temperature, where very little is known about the spectral shape, MEM
would provide new insight in the thermal changes of hadronic
properties. Moreover, this approach requires no a priori assumptions or
specific ansaetze and thus allows a reconstruction of the spectral functions
from first principles. A first application of MEM at finite temperature could
show that a reconstruction of the continuum part of the free thermal meson
spectral function is possible \cite{Wet00}. In the following a similar test is
reported for free thermal meson correlators calculated on the lattice. On the
basis of such knowledge the MEM analysis is extended to meson correlation
functions at several temperatures below and above the deconfinement transition.
\section{APPLICABILITY OF MEM AT\\FINITE TEMPERATURE}
\begin{figure}[t]
\bc
\epsfig{file=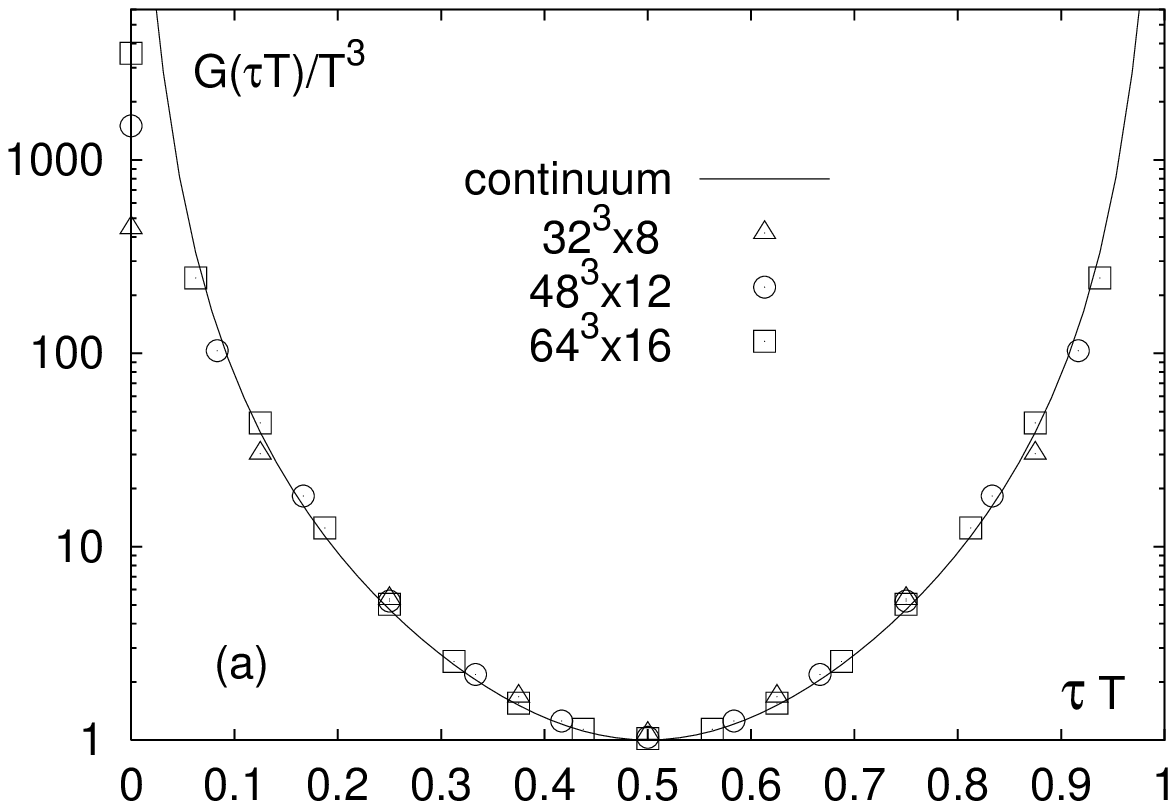,width=70.8mm}
\epsfig{file=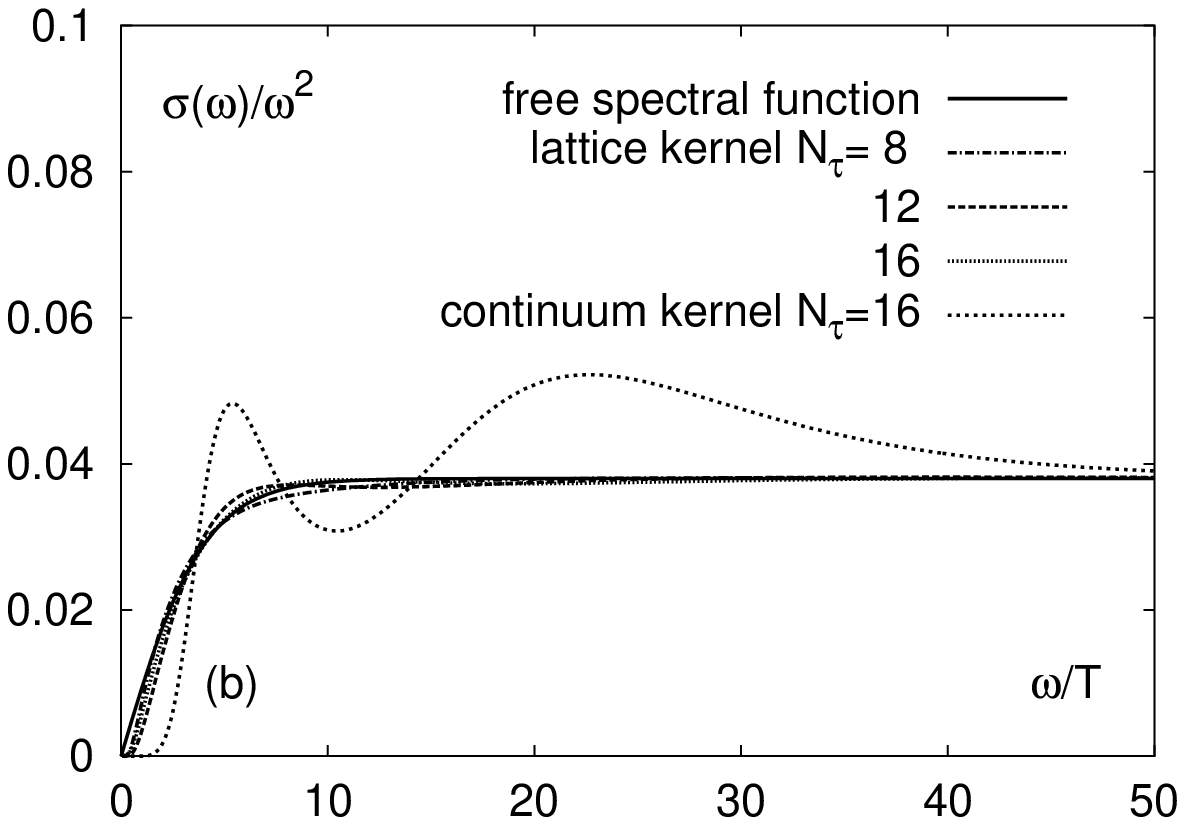,width=70.0mm}
\vskip-14mm
\ec
\caption{Thermal correlators (a) and reconstructed spectral functions
(b) with integral kernel $K$ and $K_L$}
\label{free}
\end{figure}

Instead of sharp ground state peaks at $T=0$ broad bumps and continuum-like
structures are expected to dominate the spectral shape at sufficiently high
temperature. In order to test the applicability of MEM at finite temperature we
have calculated free thermal meson correlation functions $G(\ta)=\int d^3x \;
\langle J(\ta,\vec{x}) J^\dag(0,\vec{0}) \rangle$ in the (pseudo-)scalar
channel for different lattice sizes projected to zero momentum. A Gaussian
noise with the variance $\Delta(\ta) \sim \ta \: G(\ta)$ was added to the
exact values to simulate the statistical error for correlators at finite
temperature \cite{Asa00,Wet00}. The correlator is then related to the
respective spectral function $\si(\om)$ through the integral equation
\bqa
G(\ta,T) = \int_0^\infty d\om \; \si(\om) \; K(\ta,\om) \; ,\nn
\eqa
where the continuum integral kernel is given by
\bqa
K(\ta,\om) = \frac{\cosh(\om(\ta-1/2T))}{\sinh(\om/2T)}.\nn
\eqa
The thermal correlators and reconstructed spectral functions are illustrated in
figure \ref{free}.
Strong finite size effects are visible for the correlators at small time
separations compared to the continuum curve (solid line). In the reconstructed
spectral function this results in unphysical bumps (dotted curve).

The application of a lattice adapted kernel
\bqa
K_L(\ta,\om) = {2\omega \over T} \sum_{n=0}^{N_\tau -1}
{\exp (- i 2\pi n\tau T) \over
(2 N_\tau \sin(n\pi /N_\tau ))^2 + (\omega/T)^2}\nn
\eqa
is absolutely mandatory to obtain the correct spectral shape. It can be
observed in figure \ref{free} that this choice of the kernel combined with a
restriction of the $\om/T$-range up to about $4 N_\ta$  absorbs the cut-off
effects and yields an almost perfect reconstruction of the free spectral
function $\si(\om) = \frac{3}{8 \pi^2} \om ^2 \tanh(\om/4T)$ for
$N_\ta =8, 12$ and 16.
\subsection{Smearing at finite Temperature}
\begin{figure}[t]
\bc
\epsfig{file=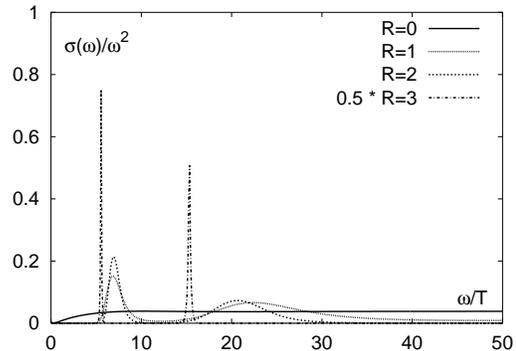,width=70.8mm}
\vskip-14mm
\ec
\caption{Influence of fuzzing \cite{UKQCD} on the thermal
spectral function for different fuzzing radii $R$}
\label{fuzz}
\end{figure}

At zero temperature the application of fuzzing \cite{UKQCD} and/or smearing
techniques to optimize the projection onto the ground state
leads to a reduction of the peak width in the spectral function and an almost
complete elimination of excited states. Therefore one might be
tempted to use this method also at finite temperature, where the peaks are less
pronounced compared to the continuum. However, the entire concept of smearing
has to fail when a single state is not well separated from higher excited
states. The situation will naturally be even worse when only a continuum
exists. In the case $T > T_c$ one can no longer be sure to project on an actual
ground state, since even the fuzzing of the free meson correlation function
leads to sharp peaks instead of the broad continuum. This is illustrated in
figure \ref{fuzz} for different fuzzing radii R. In order not to be biased in
the MEM analysis we therefore use unmodified point-point correlators only,
which preserve the full information about ground and excited states as well as
the continuum contribution.
\section{MESON SPECTRAL FUNCTIONS}
Having tested the applicability of MEM at finite temperature, we use the
approach to analyze meson spectral functions in the temperature range
0.4 to 3 $T_c$. Gauge field configurations on lattice sizes $(24-64)^3\times
16$ were generated with the plaquette gauge action, while the Clover action
with non-perturbatively improved coefficients \cite{Lue97} was used in the
fermion sector. Temporal and spatial meson correlators were calculated for four
quark mass values below $T_c$ and at almost zero quark mass (in the vicinity
of $\ka_c$) above $T_c$.

\begin{figure}[t]
\bc
\epsfig{file=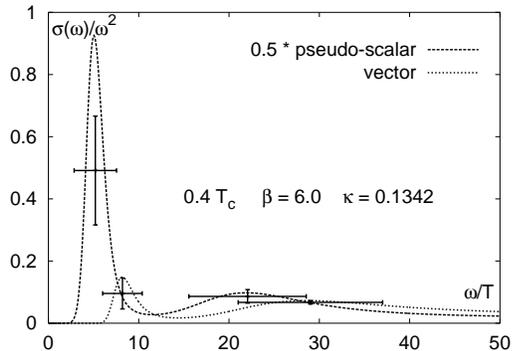,width=70.8mm}
\vskip-12.5mm
\ec
\caption{Meson spectral functions at 0.4 $T_c$}
\label{0.4Tc}
\end{figure}

In general a broadening of the peaks and a reduction in height can be observed
at finite temperature due to the short extent in the temporal direction.
Nevertheless, a good agreement of the meson ground state masses is obtained
comparing the MEM results at 0.4 $T_c$ with zero temperature data ($24^3 \times
32$ lattice \cite{Goe98}) and conventional exponential fits results of the
spatial correlator. The advantage of MEM is most obvious for the temporal
vector meson correlator, where the exponential fits lead to overestimated mass
values. MEM allows a distinction between the pole and continuum contributions
(see fig.~\ref{0.4Tc}) and thus detects precisely the actual ground state mass.
Error bars are indicated for the average spectral functions in the given
$\om/T$-range.
While the situation at\linebreak
0.4 $T_c$ and 0.6 $T_c$ is similar and almost no thermal effects are visible,
the situation is clearly different at 0.9 $T_c$. Here we observe a broadening
of the spectral functions as well as a shift in the location of the peaks.
To what extent this is a physical or statistical effect has to be clarified
with larger statistics.

\begin{figure}[t]
\bc
\epsfig{file=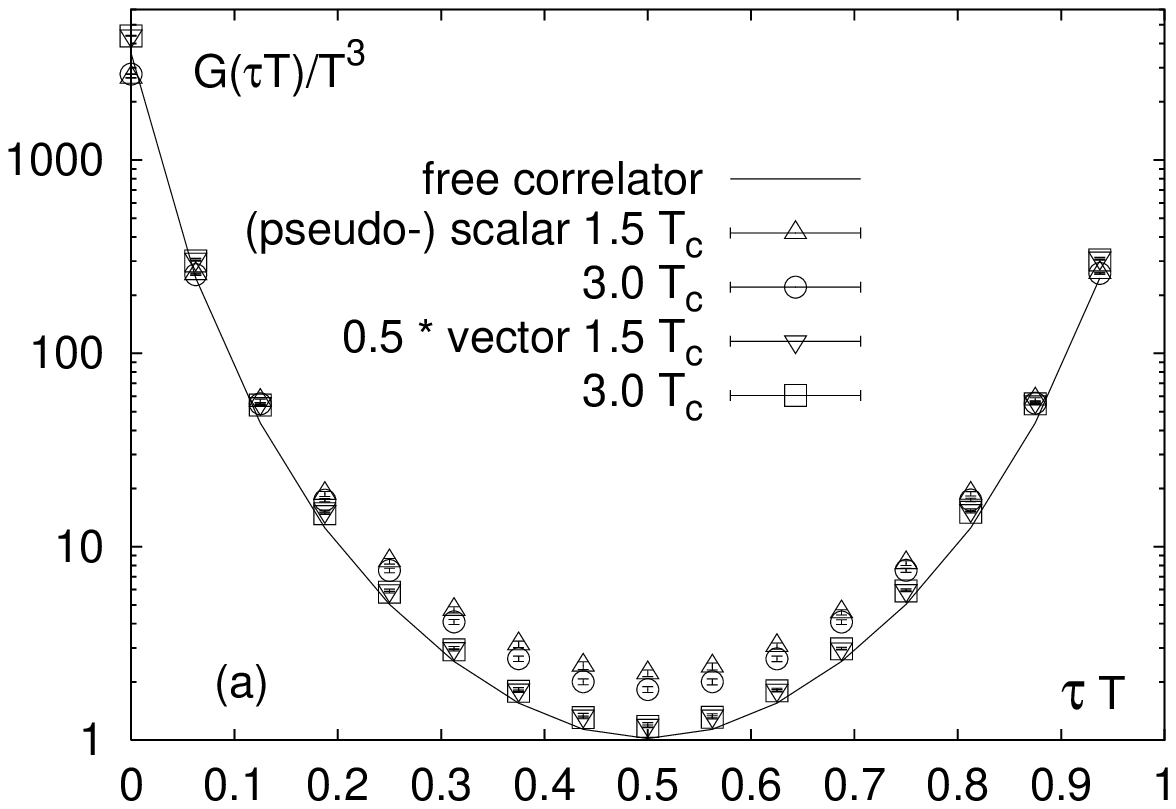,width=70.8mm}
\hspace*{-3mm}
\epsfig{file=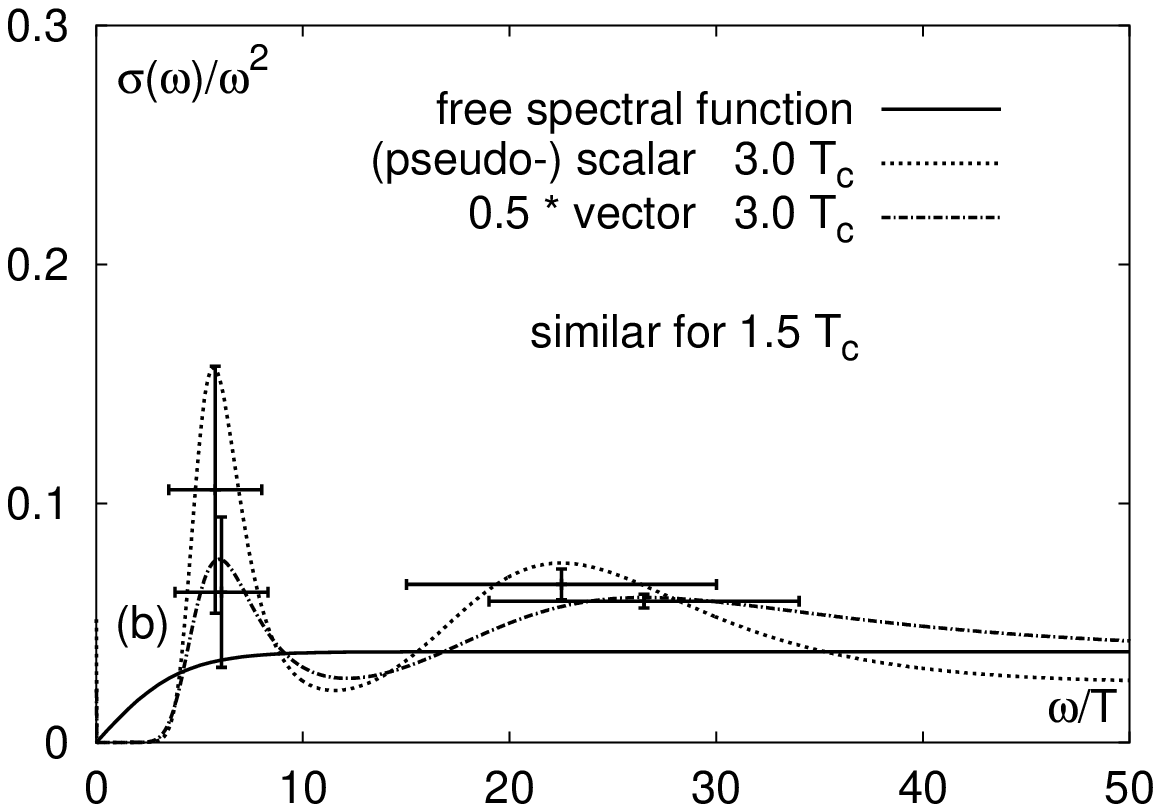,width=70.0mm}
\vskip-14mm
\ec
\caption{Meson correlators (a) and reconstructed spectral functions
(b) above $T_c$\vskip-1mm}
\label{aboveTc}
\end{figure}
Increasing the temperature above $T_c$ a gradual approach towards the free
curve is found in the (pseudo-)scalar channel (see fig.~\ref{aboveTc}a).
The vector meson correlator is instead much closer to free quark behavior.
In this channel we have summed up the contributions with all four
$\ga$-matrices, which yields twice the (pseudo-)scalar correlator in the free
case. These properties are similarly evident from the reconstructed
spectral shape in figure \ref{aboveTc}b. However, the vector meson spectral
function shows an enhancement around $\om/T\sim6$, while a suppression compared
to the free curve is found at smaller energies. This observation is in contrast
to HTL-resummed perturbation theory \cite{Kar01}, where the spectral function
is IR-divergent. The deviations in the (pseudo-)scalar channel cannot be
sufficiently explained by HTL medium effects like thermal quark masses and
Landau damping. This leads to the conclusion that the behavior in the
deconfined plasma phase is characterized by strongly correlated quarks
and gluons in the temperature range up to 3 $T_c$.
\end{document}